\documentclass[12pt]{article}
\usepackage{amsmath}
\usepackage{amssymb}
\begin{document}
\begin{center}
{\bf {\Large   {(Anti-)Chiral Superfield Approach to  Interacting Abelian 1-Form Gauge Theories:
  Nilpotent and Absolutely Anticommuting Charges}}}
  \vskip 3.0cm

{\sf         B. Chauhan$^{(a)}$, S. Kumar$^{(a)}$,  R. P. Malik$^{(a,b)}$}\\
$^{(a)}$ {\it Physics Department, CAS, Institute of Science,}\\
{\it Banaras Hindu University, Varanasi - 221 005, (U.P.), India}\\

\vskip 0.1cm

\vskip 0.1cm

$^{(b)}$ {\it DST Centre for Interdisciplinary Mathematical Sciences,}\\
{\it Institute of Science, Banaras Hindu University, Varanasi - 221 005, India}\\
{\small {\sf {e-mails: bchauhan501@gmail.com; sunil.bhu93@gmail.com; rpmalik1995@gmail.com  }}}

\end{center}

\vskip 2cm

\noindent
{\bf Abstract:}
We derive the off-shell nilpotent (fermionic) (anti-)BRST symmetry transformations
by exploiting the (anti-)chiral superfield approach (ACSA) to Becchi-Rouet-Stora-Tyutin (BRST)
formalism for the {\it interacting}  Abelian 1-form gauge theories  where there is a coupling 
between the  $U(1)$ Abelian 1-form gauge field
and Dirac as well as complex scalar fields. We exploit the (anti-)BRST invariant
 restrictions on the (anti-)chiral superfields to 
derive the {\it fermionic} symmetries of our present D-dimensional Abelian 1-form gauge  theories.
The {\it novel} observation of our present investigation is the derivation of 
the absolute anticommutativity
of the nilpotent (anti-)BRST charges {\it despite} the fact that our ordinary D-dimensional
interacting Abelian 1-form gauge theories are generalized onto the 
(D, 1)-dimensional (anti-)chiral super submanifolds (of the general (D, 2)-dimensional supermanifold)  where {\it only} 
 the (anti-)chiral super expansions of the (anti-)chiral superfields have been taken into account. 
 We also discuss the nilpotency of the (anti-)BRST charges and (anti-)BRST invariance of the 
 Lagrangian densities  of our present interacting Abelian 1-form gauge theories within the framework of ACSA to BRST formalism.
\vskip 2.0cm


\vskip 0.5cm


\newpage
\section{Introduction}

The usual superfield approach (USFA) to Becchi-Rouet-Stora-Tyutin (BRST) formalism exploits 
the idea of horizontality condition (HC) for the derivations of the 
(anti-) BRST symmetries for the guage and corresponding (anti-)ghost
 fields as well as the Curci-Ferrari condition of a non-Abelian 1-form gauge theory (see, e.g. [1-8]).
However, it does not shed any light on the (anti-)BRST symmetry transformations
associated with the {\it matter} fields of an {\it interacting} (non-)Abelian 
gauge theory where there is  a coupling between the gauge and matter fields.
The USFA has been systematically generalized so as to derive the nilpotent (anti-)BRST 
symmetry transformations for the {\it matter}, gauge  and (anti-)ghost 
fields {\it together}. The latter superfield approach has been christened
as the augmented version of superfield approach (AVSA) to BRST formalism
(see, e.g. [9-12]). In the above superfield approaches [1-12], the {\it full} super
expansions of the superfields have been taken into account along {\it all}
possible Grassmannian directions of the (D, 2)-dimensional supermanifold on which the ordinary
D-dimensional
gauge theory is generalized.

In our recent works [13-15], a simpler version of the AVSA to BRST formalism has been proposed  
where {\it only} the (anti-)chiral superfields have been taken into account 
for the derivation of the (anti-)BRST symmetry transformations. In the USFA/AVSA to
BRST formalism, a given  D-dimensional gauge theory is generalized onto 
a (D, 2)-dimensional supermanifold (characterized by the superspace coordinates 
$Z^M = ( x^{\mu}, \theta, \bar\theta)$ where $x^{\mu}\;(\mu = 0, 1...D-1)$
are the bosonic coordinates and a pair of Grassmannian variables $(\theta, \bar\theta)$
satisfy: $\theta^2 = 0 =\bar\theta^2,\; \theta\bar\theta + \bar\theta\theta = 0$).
On the contrary, in the (anti-)chiral superfield approach (ACSA) to BRST
formalism, a given D-dimensional gauge theory is generalized onto the
(D, 1)-dimensional (anti-)chiral super-submanifolds of the {\it general}
 (D, 2)-dimensional  supermanifold. In our present endeavor, we have 
 considered ACSA to BRST formalism and discussed various aspects 
of the (anti-)BRST symmetries and (anti-)BRST charges of an interacting 
Abelian 1-form gauge theory where there is a coupling between the $U(1)$
gauge field $(A_\mu)$ and fermionic $(\psi^2 =\bar\psi^2 =0,
\psi\bar\psi + \bar\psi\psi$) Dirac fields $(\bar\psi,\;\psi)$. We have {\it also} discussed
briefly the interacting Abelian 1-form $U(1)$ gauge theory  with complex scalar fields
 where there is a coupling between the gauge  and matter fields 
(i.e. complex scalar fields) and
have utilized the potential and power of the ACSA to BRST formalism 
to derive the proper (anti-)BRST  symmetries (cf. Appendix A).

The key results of our present investigation are the proof of  nilpotency
and absolute anticommutativity of the (anti-)BRST conserved charges within
the framework of ACSA to BRST formalism. The derivation of the (anti-)BRST
symmetries and their nilpotency properties have been discussed in {\it all} the previous
works [9-15]. However, the proof of the absolute anticommutativity of the (anti-)BRST charges
is a {\it novel} result because, in our earlier works on AVSA, we have {\it not} discussed this
aspect of the (anti-)BRST charges [9-12]. In fact, the discussion about the ACSA to BRST
formalism and the proof of absolute anticommutativity of the   nilpotent (anti-)BRST charges
(within the framework of ACSA to BRST formalism) have been a set of challenging problems for us
and we have resolved these issues in our present investigation in an elegant manner ({\it despite}  the fact
that we have taken into account  {\it only} the (anti-)chiral super expansions of the superfields).
We have derived the (anti-)BRST symmetries by imposing  the (anti-)BRST invariant restrictions
on the (anti-)chiral superfields which are the {\it quantum} analogues of the {\it classical}
gauge invariant restrictions (GIRs) that have been utilized in our earlier works [9-12].

Against the backdrop of the above discussions, it is pertinent to point out that the
ACSA has  {\it also} been  applied to  $\mathcal{N} = 2$  supersymmetric (SUSY) quantum mechanical systems
of interest in our earlier works [16-19] where we have derived the nilpotent   $\mathcal{N} = 2$ 
SUSY transformations in an elegant manner. We have also derived the conserved  $\mathcal{N} = 2$ 
SUSY charges and expressed them in terms of the supervariables obtained after the appropriate 
 $\mathcal{N} = 2$  SUSY invariant restrictions. However, these charges do {\it not}
obey the absolute anticommutativity property\footnote{In the case of $\mathcal{N} = 2$ SUSY
quantum mechanical models, the anticommutator of two distinct  SUSY transformations 
on a {\it variable} leads to the time derivative on that {\it specific}  variable.}. We have been able to capture the nilpotency 
property  of the super charges in the terminology of the (anti-)chiral supervariables.
 We have {\it not} been  able to say anything, however, about the absolute anticommutativity
property   between two  $\mathcal{N} = 2$  SUSY  conserved charges. Thus, in our present endeavor, the proof of the absolute anticommutativity 
property between the  (anti-)BRST charges  is a completely {\it novel} result.

The main motivations behind our present endeavor are as follows.
First of all, we demonstrate that the absolute anticommutativity
of the (anti-)BRST charges is {\it true} even if we take {\it only}
the (anti-)chiral super expansions of the superfields.
This is a novel observation within the framework of ACSA to BRST
formalism. Second, we have established a surprising observation 
that the anticommutativity of the BRST charge with anti-BRST charge is 
deeply connected with the nilpotency ($\partial_{\theta}^2 = 0$)
property of the translational generator ($\partial_{\theta}$) along
the $\theta$-direction of the {\it chiral} super submanifold (and the anticommutativity
of the anti-BRST charge with BRST charge is intimately  related to the nilpotency
($\partial_{\bar\theta}^2 = 0$) of the translational generator ($\partial_{\bar\theta}$)
 along  $\bar\theta$-direction of the {\it anti-chiral} super submanifold).
Third, our present idea has been generalized to the proof of 
nilpotency and absolute anticommutativity of the (anti-)BRST 
charges of an {\it interacting} $SU(N)$ non-Abelian gauge theory [20], too.
Finally, our method of derivation supports the results that have 
been obtained from the mathematically precise use of HC for the self-interacting (non-)Abelian
1-form theory {\it without}  any interaction with matter fields (see, e.g. [4, 5] for details).

Theoretical material of our present paper is organized as follows.
In Sec. 2, we discuss the bare essentials of (anti-)BRST symmetries 
for the Lagrangian density  of an {\it interacting} D-dimensional Abelian 1-form gauge
theory in the Feynmen gauge and derive the conserved charges.
Our Sec. 3 deals with the ACSA to BRST formalism where we derive the
BRST symmetries using the {\it anti-chiral} superfields.
Sec. 4 of our present endeavor is devoted to the derivation of anti-BRST symmetries by 
using the ACSA to BRST formalism where the {\it chiral} superfields are utilized.
In Sec. 5, we express the conserved (anti-)BRST charges on the (D, 1)-dimensional 
super submanifolds (of the general (D, 2)-dimensional supermanifold on which our
theory is generalized) and provide  
 the proof of nilpotency and absolute anticommutativity properties of the (anti-)BRST
charges within the framework of ACSA to BRST formalism.
We discuss the (anti-)BRST invariance of the Lagrangian density, within the framework of 
ACSA to BRST formalism, in Sec. 6.
 Finally, we summarize our key
results in Sec. 7 and point out a few theoretical directions for future investigations
within the framework of superfield formalism.

In our Appendix A, we discuss  the  absolutely anticommuting   (anti-)BRST
charges for the {\it interacting} Abelian 1-form gauge theory  where there is a coupling between the
$U(1)$ gauge field and complex scalar fields. The subject matter of Appendix B concerns itself with the natural
and automatic proof of the absolute anticommutativity 
property of the (anti-)BRST  symmetries  (and corresponding  charges) 
when the {\it full} super expansions of the
superfields are taken into account.

\section
{\bf Preliminaries: Lagrangian Formulation}
First of all, we begin with the {\it interacting} D-dimensional Abelian 1-form gauge theory where there  is a
coupling between the $U(1)$ gauge field $(A_\mu)$ and the Dirac fields ($\bar\psi, \psi$).
The Lagrangian density for this system, in the Feynmen gauge, is as follows\footnote{The 
background flat D-dimensional Minkowskian spacetime manifold is endowed with a metric tensor
$\eta_{\mu\nu}$ = diag $(+1, -1, -1, ....)$. This implies that the short-hand notations,
in their explicit forms, are:
 $(\partial\cdot A) = \eta _{\mu\nu}\partial^{\mu} A^{\nu} \equiv 
\partial_{\mu} A^\mu\equiv\partial_0\;A_0-\partial_i A_i$ and $\square = \eta_{\mu\nu}\partial^\mu\partial^\nu = \partial_0^2- \partial_i^2$.
Here the Greek indices $\mu, \nu, \lambda... = 0, 1, 2... D-1$ and Latin indices  $ i, j, k...= 1, 2...D-1 $ correspond to the 
spacetime and space directions, respectively, on the flat Mikowskian spacetime manifold.}
 (see, e.g. [21])
\begin{eqnarray}
&&{\cal L}_B = -\,\frac{1}{4}F^{\mu\nu}F_{\mu\nu} + \bar\psi\,(i\,\gamma ^ \mu D_\mu - m)\,\psi + B\,(\partial\cdot A)
+ \frac{B^2}{2} - i\,\partial_{\mu}\bar C\, \partial^{\mu} C,
\end{eqnarray} 
where $F_{\mu\nu} = \partial_\mu A_\nu - \partial_\nu A_\mu $ is the  field strength tensor  and $D_\mu\psi = \partial_{\mu}\psi +
 i\, e\, A{_\mu}\,\psi $ is the covariant derivative on the Dirac field $\psi$. The gauge-fixing term
$(-\,\frac{1}{2} (\partial\cdot A)^2) $ has been linearized by invoking 
the Nakanishi-Lautrup auxiliary field $B$, $\gamma^\mu$ (with $ \mu = 0, 1, 2...D-1$) are the  ($D\times D)$ Dirac
gamma matrices, $m$ is the rest mass of the  Dirac particle  and the fermionic ($C^2 = \bar C^2 = 0, \, C\,\bar C + \bar C\, C = 0$)
(anti-)ghost fields $(\bar C) C$ are needed for the validity of unitary in the theory.
It is evident that the fermionic fields ($\psi,\,\bar\psi,\, C,\, \bar C$) anticommute among themselves 
and they commute with the bosonic fields $A_\mu$ and $B$ of our theory. It is also elementary 
to state that the bosonic fields commute among themselves.

The above Lagrangian density respects the following infinitesimal, continuous,  off-shell nilpotent ($s_{(a)b}^2 = 0)$
and absolutely anticommuting ($s_b s_{ab} + s_{ab} s_b = 0$) (anti-)BRST transformations $s_{(a)b}$
\begin{eqnarray}
&&s_{ab} A_\mu = \partial_\mu \bar C,\qquad\;\; s_{ab} \bar C = 0,\qquad\qquad\;\; s_{ab} C = -\,i\,B,\qquad s_{ab} B = 0,\qquad\nonumber\\
&&s_{ab} \bar\psi = -\,i\,e\,\bar\psi \,\bar C,\quad\; s_{ab} \psi = -\,i\,e\,\bar C\;\psi,\quad\, s_{ab }F_{\mu\nu} = 0,\quad s_{ab}(\partial\cdot A) = \square\bar C, \nonumber\\
&&s_b A_\mu = \partial_\mu C,\qquad\quad s_b C = 0,\qquad\qquad\;\;\; s_b \bar C = i\,B,\qquad\quad\;\,  s_b B = 0,\nonumber\\
&&s_b \bar\psi = -\,i\,e\,\bar\psi\, C,\quad\;\; s_b \psi = -\,i\,e\,C\, \psi,\quad\;\;\, s_b F_{\mu\nu} = 0,\quad s_b (\partial\cdot A) = \square C,
\end{eqnarray} 
because the Lagrangian density (1) transforms to the total spacetime derivatives 
\begin{eqnarray}
&&s_{ab}\;{\cal L}_ B  = \partial_\mu [B\,\partial^\mu \bar C],\qquad\qquad s_b\;{\cal L}_B  = \partial_\mu [B\,\partial^\mu C],
\end{eqnarray}
thereby rendering the action integral $S = \int d{^D}x\;{\cal L}_B$ invariant for the  physically well-defined fields that
vanish off at infinity. The conserved currents, due to Noether's theorem, are:
\begin{eqnarray}
&&J^{\mu}_{ab} = -\,F^{\mu\nu}\partial_{\nu}\bar C + B\;\partial^\mu \bar C - e\,\bar\psi\;\gamma^\mu \;\bar C \;\psi,\nonumber\\ 
&&J^{\mu}_b = -\,F^{\mu\nu}\partial_{\nu}C + B\;\partial^\mu C - e\,\bar\psi\;\gamma^\mu \;C \;\psi.  
\end{eqnarray}
Using the following Euler-Lagrange (EL) equations of motion (EOM)
\begin{eqnarray}
&&\partial_\mu F^{\mu\nu} - \partial^\nu B = e\,\bar\psi\,\gamma^\nu\, \psi, \qquad \Box C = 0,\qquad 
(i\,\gamma ^ \mu \partial_\mu - m)\,\psi = e \;\gamma^{\mu}\; A_{\mu}\;\psi ,\nonumber\\
&&i\,(\partial_\mu\bar\psi)\,\gamma^\mu + m\bar\psi = -\,e\;\bar\psi\;\gamma^\mu \;A_\mu,\qquad B = -\;(\partial\cdot A),\qquad\Box\bar C = 0,
\end{eqnarray}
it can be readily checked that $\partial_{\mu}J_{(a)b}^{\mu} = 0$. Thus, the conserved and the off-shell nilpotent
 (anti-)BRST charges $Q_{(a)b} = \int d^{D-1}x \;J_{(a)b}^0 $ 
can be expressed  (with $\dot B = \frac {\partial B}{\partial t}, \dot C = \frac {\partial C}{\partial t},$ etc.) as follows
\begin{eqnarray}
&&Q_{ab} =   \int d^{D-1}x\;[-\,F^{0i}\partial_i \bar C + B\, \dot{\bar C} - e\,\bar\psi\,\gamma^0 \bar C\,\psi]\; 
\equiv \int d^{D-1}x\;[B\, \dot{\bar C} - {\dot B} \bar C],
\end{eqnarray}
\begin{eqnarray}
&&Q_b =  \int d^{D-1}x\;[-\,F^{0i}\partial_i  C + B\, {\dot C} - e\,\bar\psi\,\gamma^0  C\,\psi]\; 
\equiv  \int d^{D-1}x\;[B\, \dot{C} - {\dot B}  C],
\end{eqnarray}
where we have used the EOM: $\partial_i F^{0i} = -\,(\dot B + e\,\bar \psi\,\gamma^0\,\psi)$ and carried out a partial
 integration to drop the total space derivative term due to Gauss's divergence theorem. We have also used the 
 convention of left-derivative while deriving the EL-EOM w.r.t. to the  fermionic fields $(C, \bar C, \psi, \bar\psi)$
 of our present {\it interacting} D-dimensional  Abelian 1-form gauge theory with Dirac's fields $(\psi, \bar\psi)$.

The above conserved charges are nilpotent $(Q_{(a)b}^2 = 0)$ of order
 two and absolutely anticommuting ($Q_b Q_{ab} + Q_{ab} Q_b = 0$) in nature (cf. Sec. 5 below for details).
In fact, the above conserved charges $(Q_{(a)b})$ in (6) and (7) are
 the generators of the infinitesimal, continuous and nilpotent (anti-)BRST  symmetry transformations 
$s_{(a)b}$. In other words, we have the following explicit relationship 
\begin{eqnarray}
s_{(a)b}\;\Phi  = \mp\,i\,[\Phi, Q_{(a)b}]_{\pm }, 
\end{eqnarray}
where $\Phi$($ = A_\mu, C, \bar C, B, \psi, \bar\psi$) is the generic field of the theory and ($\pm$) signs, as the subscripts on the square bracket, 
denote the (anti)commutator for the generic  field $\Phi$ being fermionic/bosonic in nature. The $(\mp)$ signs in front of the 
bracket has to be chosen [22] judiciously  for the derivation of nilpotent and absolutely anticommuting
(anti-)BRST symmetry transformations (i.e. $s_{(a)b}\; \Phi$).

\section
{\bf Nilpotent BRST Symmetries: Anti-Chiral Superfields and Their Super Expansions}

We derive here the nilpotent BRST symmetries of Eq. (2) by applying ACSA to BRST formalism
 where we use the anti-chiral superfields {\it only}. To this end in mind, first of all, we generalize
the {\it ordinary} fields of the Lagrangian density (1) onto (D, 1)-dimensional anti-chiral super-submanifold 
(of the {\it general} (D, 2)-dimensional supermanifold) as follows:
\begin{eqnarray}
 &&A_{\mu} (x) \longrightarrow  B_\mu (x, \bar\theta) = A_{\mu}(x) + \bar\theta\, R_\mu (x),\quad
 C (x) \longrightarrow F(x, \bar\theta) = C(x) + i\,\bar\theta\, B_{1}(x),\nonumber\\
&& \bar C (x) \longrightarrow \bar F(x, \bar\theta)~~~ = \bar C_(x) + i\,\bar\theta\; B_{2} (x),\quad
 \psi (x)  \longrightarrow \Psi  (x, \bar\theta) = \psi (x) + i\,\bar\theta\, b_1 (x),\nonumber\\
&& \bar\psi (x)  \longrightarrow \bar\Psi  (x, \bar\theta)~~~ = \bar\psi (x) + i\,\bar\theta\, b_2 (x),\quad
 B (x)  \longrightarrow \tilde B (x, \bar\theta) = B(x) + i\,\bar\theta\, f(x), 
 \end{eqnarray} 
where the fields  ($ R_\mu $, $f$) are the fermionic secondary fields and ($B_1, B_2, b_1, b_2$) are the bosonic
secondary fields that have to be determined in terms of the basic and auxiliary fields of the theory (cf. Eq. (1))
by invoking the
BRST invariant restrictions\footnote{The BRST and anti-BRST invariant restrictions are 
at the {\it quantum} level and these are the analogues of the {\it classical} gauge invariant restrictions (GIRs)
where we demand that the {\it physical} (i.e. gauge invariant) quantities should be independent of the ``soul" coordinates $(\theta, \bar\theta)$.}.
It is straightforward to note that the ($D$, 1)-dimensional anti-chiral super-submanifold is parameterized by $x^\mu$ and $\bar\theta$.
This is why, it is called as anti-chiral.

According to the basic tenets of ACSA/AVSA to BRST formalism, the BRST invariant quantiles {\it must} remain 
independent of the Grassmannian variable ($\bar\theta$) when they are generalized onto the (D, 1)-dimensional 
anti-chiral super submanifold (see, e.g. [9-15]). Such useful and interesting  BRST invariant quantities are:
\begin{eqnarray}
&& s_b C = 0, \qquad s_b B = 0,\qquad s_b (\bar\psi\;\psi) = 0,\qquad s_b (\bar\psi D_\mu\psi) = 0,\qquad s_b (A^{\mu}\;\partial_\mu C) = 0, \nonumber\\
&& s_b[A^{\mu}\;\partial_\mu B + i\;\partial_\mu \bar C\; \partial^\mu C] = 0,\qquad\quad s_b (C\;\psi) = 0,\qquad s_b (\bar\psi\; C) = 0.
\end{eqnarray}
The above quantities are to be generalized onto the (D, 1)-dimensional anti-chiral super-submanifold with 
the following restrictions:
\begin{eqnarray}
&&F(x, \bar\theta) = C(x), \;  \;\bar\Psi (x, \bar\theta)\; F(x, \bar\theta) = \bar\psi (x)\; C(x),\quad\bar\Psi(x,\bar\theta)\;\Psi(x, \bar\theta) = \bar\psi (x)\;\psi(x)\nonumber\\
&&B^\mu (x, \bar\theta)\;\partial_\mu F(x, \bar\theta) = A^\mu (x) \partial_\mu C(x),\;\; F(x, \bar\theta)\;\Psi(x, \bar\theta) = C(x)\;\psi(x),\nonumber\\
&&\tilde B (x, \bar\theta) = B(x),\quad \bar\Psi (x, \bar\theta)\;\partial_\mu \Psi(x, \bar\theta) + i\,e\;\bar\Psi (x, \bar\theta)\;B_\mu (x, \bar\theta)\Psi(x, \bar\theta)\nonumber\\
&& = \bar\psi(x)\partial_\mu \psi (x) + i\,e\;\bar\psi (x)\;A_\mu (x)\;\psi (x),\;\, B^\mu (x, \bar\theta)\; \partial_\mu\tilde B(x, \bar\theta) + i\;\partial_{\mu}\;\bar F(x, \bar\theta)\;\partial^{\mu}F(x, \bar\theta)\nonumber\\
&& = A^{\mu}(x)\;\partial_{\mu}B(x) + i\; \partial_{\mu}\bar C(x)\;\partial^{\mu}C(x).
\end{eqnarray}
These restrictions lead to the derivation of the  expressions for the secondary fields (of  
the super expansions  (9)) in terms of the basic and auxiliary fields as:
\begin{eqnarray}
&&R_\mu = \partial_\mu C, \qquad B_1 (x) = 0, \qquad B_2 (x) = B(x), \nonumber\\
&& b_1 = -\,e\,C\,\psi,\qquad b_2  = -\,e\,\bar\psi\; C,\qquad f(x) = 0.
\end{eqnarray}
We elaborate on a few of the above derivations here. It is evident that the first entry and the restriction 
$\tilde B (x, \bar\theta) = B(x)$ in (11)
produce the following results (with $f(x) = 0, B_1(x) = 0$), namely;
\begin{eqnarray}
F^{(b)}(x, \bar\theta) \,&=& \, C(x) + \bar\theta\,(0) \quad\equiv\quad C(x) + \bar\theta\,(s_b \,C(x)),\nonumber\\
\tilde B^{(b)}(x, \bar\theta)\, &=& \, B(x) + \bar\theta\,(0) \quad\equiv\quad B(x) + \bar\theta\,(s_b\, B(x)),
\end{eqnarray}
where the superscript $(b)$ stands for the super expansions of the  anti-chiral superfields that have been derived after the BRST
invariant restrictions (11) and which
lead to: $s_b\;C = 0, s_b\; B = 0$. We use (13) now in 
\begin{eqnarray}
&&B^\mu (x, \bar\theta)\;\partial_\mu F^{(b)}(x, \bar\theta) = A^\mu (x)\; \partial_\mu C(x),\nonumber\\
&&B^\mu (x, \bar\theta) \;\partial_\mu\tilde B^{(b)}(x, \bar\theta) + i\;\partial_{\mu}\bar F(x, \bar\theta)\;\partial^{\mu}F^{(b)}(x, \bar\theta)\nonumber\\
&& =
A^{\mu}(x)\;\partial_{\mu}B(x) + i\; \partial_{\mu}\bar C(x)\;\partial^{\mu}C(x),
\end{eqnarray}
leading to the following relationships:
\begin{eqnarray}
R^\mu (x) \partial_\mu C(x) &=& 0 \quad \Longrightarrow \quad R_\mu (x)\propto \partial_\mu C(x),\nonumber\\
R^\mu (x) \partial_\mu B(x) - \partial_\mu B_2 (x) \partial^\mu C(x) &=& 0\quad \Longrightarrow \;\, R_\mu = \partial_\mu C(x),\,\, B_2 (x) = B(x).
\end{eqnarray}
We have chosen, for the shake of brevity: $R_\mu = \partial_\mu C$, which implies that $B_2 (x) = B(x)$. These inputs lead 
to the following: 
\begin{eqnarray}
&&B_{\mu}^{(b)}(x, \bar\theta) = A_\mu (x) + \bar\theta\, (\partial_\mu C(x))\,\equiv \,A_\mu (x) + \bar\theta \,(s_b A_\mu (x)),\nonumber\\
&& \bar F^{(b)}(x, \bar\theta) = \bar C(x) + \bar\theta \,(i\,B(x))\,\equiv \,\bar C(x) + \bar\theta\, (s_b \bar C (x)).
\end{eqnarray}
Thus far, we have derived the BRST symmetry transformations for the gauge field and corresponding (anti-)ghost fields
of our present D-dimensional {\it interacting} Abelian theory with Dirac's fields which are usually derived by exploiting the theoretical
power and potential of HC (see, e.g. [4, 5]).

We are in the position  now to derive the BRST symmetry transformations for the matter fields ($\bar\psi,\psi $).
Towards this goal in mind, we note that the restrictions, corresponding to the BRST invariances   $s_b (C\;\psi) = 0$ and  $s_b (\bar\psi\; C) = 0$ 
 in (11), imply: $ C(x)\;b_1 (x) = 0$ and $b_2 (x)\; C(x) = 0$. 
In other words, the secondary fields $b_1 (x)$ and $b_2 (x)$ are {\it proportional} to the ghost field $C(x)$.
The condition $\bar\Psi (x, \bar\theta) \,\Psi(x, \bar\theta) = \bar\psi(x) \, \psi(x)$ leads us to conclude that $b_2 \;\psi = \bar\psi\; b_1 $.
With these inputs, we write down the final restriction of Eq. (11) as:
\begin{eqnarray} 
&&\bar\Psi(x, \bar\theta)\;\partial_\mu \Psi (x, \bar\theta) + i\;e\;\bar\Psi (x, \bar\theta)\; B^{(b)}_{\mu}(x, \bar\theta)\;\Psi(x, \bar\theta)\nonumber\\
 &&= 
\bar\psi (x)\;\partial_\mu \psi (x) + i\;e\;\bar\psi (x)\; A_\mu (x)\; \psi (x),
\end{eqnarray}
where the exact form of $B^{(b)}_{\mu}(x, \bar\theta)$ has been illustrated in Eq. (16). The above equality,
ultimately, leads to the following:
\begin{eqnarray}
i\;e\;A_\mu \;(b_2\;\psi - \bar\psi \;b_1) - \bar\psi\;\partial_\mu b_1 + b_2 \;\partial_\mu \psi - e\;\bar\psi \;\partial_\mu C \;\psi = 0.
\end{eqnarray}
The first term is zero because $b_2 \;\psi = \bar\psi\; b_1$ which has been discussed in the paragraph above Eq. (17).
In fact, the restriction $\bar\Psi (x, \bar\theta)\;\Psi (x,\bar\theta) = \bar\psi(x)\;\psi(x)$ leads to it. The
rest of the terms (with inputs $b_1 \propto C(x)$  and $b_2 \propto C(x)$) are satisfied  by the following choices:
\begin{eqnarray}
b_1 = -\;e\;C\;\psi,\qquad b_2 = -\;e\;\bar\psi\; C\;\;\Longrightarrow \;\;b_2\; \psi  = \bar\psi \;b{_1}.
\end{eqnarray}
Thus, finally, we have obtained the following super expansions\footnote{We have taken the 
coefficient of $\bar\theta$ as the BRST transformation on an ordinary field because it
has already been proven that $s_b\longleftrightarrow \partial_{\bar\theta}$ (see, e.g.  [1-15] for details). } for {\it all} the superfields of our theory (on the anti-chiral
(D, 1)-dimensional super submanifold), namely;
\begin{eqnarray}
B_{\mu}^{(b)} (x, \bar\theta) & = & A_{\mu}(x) + \bar\theta \;(\partial_\mu C)\;\quad \equiv ~~\; A_{\mu}(x) + \bar\theta\;(s_b\; A_\mu(x)),\nonumber\\
F^{(b)}(x, \bar\theta) &=& C(x) + \bar\theta\;(0)\qquad\quad\equiv~~ C(x) + \bar\theta\ (s_b\; C(x)),\nonumber\\
\bar F^{(b)}(x, \bar\theta) &=& \bar C(x) + \bar\theta\;(i\;B)\qquad\,\equiv~~ \bar C(x) + \bar\theta\ (s_b \;\bar C(x)),\nonumber\\\
\Psi ^{(b)}(x, \bar\theta) &=& \psi (x) + \bar\theta\;(-\,i\,e\,C\,\psi )\equiv ~~\psi (x) + \bar\theta\ (s_b \;\psi(x)),\nonumber\\
\bar\Psi ^{(b)}(x, \bar\theta) &=& \bar\psi (x) + \bar\theta\;(-\,i\,e\,\bar\psi \,C)\equiv ~~ \bar\psi (x) + \bar\theta\ (s_b \;\bar\psi(x)),\nonumber\\
\tilde B^{(b)} (x, \bar\theta) &=& B(x) + \,\bar\theta\,(0) \qquad \quad\equiv ~~  B(x) +\bar \theta\; (s_b \;B(x)), 
\end{eqnarray}
where the superscript $(b)$ on the anti-chiral superfields denotes the fact that these superfields have been obtained after the use of 
BRST invariant quantities (10) (which lead to their explicit form in Eq. (11)). It is also evident that all the relationships listed in (12)
are correct and the coefficients of $\bar\theta $ in (20) are nothing but the BRST symmetry  transformations\footnote{ It is 
worth pointing out here that the mathematical power of  HC leads to the 
derivation of (anti-) BRST symmetry transformations for the basic  fields $(A_\mu, C, \bar C)$
of the theory (cf. Eqs. (13) and (16)). However, it does not shed any light on the 
(anti-)BRST  transformations for the matter fields $(\psi, \bar\psi)$ which have been derived in Eq. (20)
due to the BRST invariant restrictions (cf. Eqs. (10) and (11)) that have been considered and utilized in our present endeavor.} quoted in Eq. (2)
for the Lagrangian density (1).  Hence, we have
the mapping between the BRST symmetry transformation and partial derivative w.r.t. Grassmannian
variable on the anti-chiral super-submanifold as: $s_b\longleftrightarrow \partial_{\bar\theta}$ [1-15].

\section
{\bf  Nilpotent Anti-BRST Symmetries: Chiral Superfields and Their Super Expansions}

We derive, in this section, the nilpotent anti-BRST symmetry transformations for the Lagrangian 
density (1) which are listed in Eq. (2). To accomplish this goal precisely, we have to generalize 
the {\it ordinary} fields of Lagrangian density (1) onto the (D, 1)-dimensional {\it chiral}
super submanifold (of the {\it general} (D, 2)-dimensional supermanifold) as
\begin{eqnarray}
&& A_{\mu} (x) \longrightarrow  B_\mu (x, \theta)  =  A_{\mu}(x) + \theta\, \bar R_\mu (x),\;\;
 C (x) \longrightarrow F(x, \theta)  =  C(x) + i\,\theta\, \bar B_{1}(x),\nonumber\\
&& \bar C (x) ~~\longrightarrow \bar F(x, \theta)  =  \bar C_(x) + i\,\theta\, \bar B_{2} (x),\,\;
 \psi (x)  \longrightarrow \Psi  (x, \theta)  =  \psi (x) + i\,\theta\;\bar b_1 (x),\nonumber\\
 &&\bar\psi (x) ~~ \longrightarrow \bar\Psi  (x, \theta) = \bar\psi (x) + i\,\theta\, \bar b_2 (x),\;\,
 B (x)  \longrightarrow \tilde B (x, \theta) = B(x) + i\,\theta\, \bar f(x), 
 \end{eqnarray}
 where the pair of secondary fields ($\bar R_{\mu}, \bar f$) are fermionic in nature in  contrast to the secondary fields 
 ($\bar B_1, \bar B_2, \bar b_1, \bar b_2$) which are bosonic. These secondary fields would be determined 
 in terms of the basic and auxiliary fields of the Lagrangian density (1) by exploiting the strength of ACSA/AVSA to BRST 
formalism where the anti-BRST invariant quantities  would be required to be independent of the Grassmannian  variable 
$\theta$ (which characterizes  the {\it chiral} super-submanifold along with the bosonic coordinates $x^{\mu}$).

In the above connection, we have the following:
\begin{eqnarray}
&&s_{ab}\;\bar C = 0, \;\, s_{ab}B = 0,\;\, s_{ab}\;(\bar\psi\;\psi) = 0, \;\, s_{ab}\;(\bar\psi \,D_\mu \psi) = 0, 
\quad s_{ab}\;(A^\mu \partial_\mu \bar C) = 0,\nonumber\\
&&s_{ab}\;(\bar C\; \psi) = 0, \quad s_{ab}\;(\bar\psi\;\bar C) = 0, 
\qquad s_{ab}\;(A^\mu \,\partial_\mu B + i\;\partial_\mu \bar C\,\partial^{\mu} C) = 0.
\end{eqnarray}
Thus, the above quantities are anti-BRST invariant and, therefore, they should be independent of $\theta$ when these are generalized 
onto the {\it chiral} super-submanifold according to the basic tenets of ACSA to BRST formalism. In other words, we have the  following equalities:
\begin{eqnarray}
&&\bar F(x, \theta) = \bar C(x),\;\; \bar\Psi(x,\theta)\;\Psi(x, \theta) = \bar\psi (x)\;\psi(x),\;\;\bar\Psi(x,\bar\theta)\;\bar F(x,\bar\theta) =\bar\psi (x)\;\bar C(x),\nonumber\\
&&\bar\Psi (x, \theta)\;\partial_\mu \Psi(x, \theta) + i\,e\;\bar\Psi (x, \theta)\;B_\mu (x, \theta)\;\Psi(x, \theta) = 
\bar\psi(x)\;\partial_\mu \psi (x)\nonumber\\
&& +  i\,e\;\bar\psi (x)\;A_\mu (x)\;\psi (x), \;\;B^\mu (x, \theta)\;\partial_\mu \bar F(x, \theta) = A^\mu (x) \partial_\mu \bar C(x),\nonumber\\
 &&\bar F(x, \theta)\;\Psi(x, \theta) = \bar C(x)\;\psi(x),\;\tilde B (x, \theta) = B(x),\;\; B^{\mu}(x,\theta)\;\partial_{\mu}\tilde B(x,\theta) \nonumber\\
&& +\; i\;\partial_{\mu}\bar F(x,\theta)\; \partial^{\mu} F(x,\theta)
 = A^{\mu}(x)\,\partial_{\mu}B(x) + i\;\partial_{\mu}\bar C(x)\,\partial^{\mu}C(x).
 \end{eqnarray}
The above restrictions lead to the derivation of  secondary fields in terms of the basic and auxiliary fields
of the Lagrangian density (1) as follows:
\begin{eqnarray}
&&\bar R_\mu = \partial_\mu \bar C,\qquad \bar b_1 = -\,e\,\bar C\,\bar\psi,\qquad \bar b_2  = -\,e\;\bar\psi \;\bar C,\nonumber\\
&&\bar f(x) = 0, \qquad \bar B_1 = -\,B, \qquad \bar B_2  = 0.
\end{eqnarray}
The process of derivation is {\it same} as the {\it one} that has been adopted and used in the case of 
BRST symmetries where we have exploited the BRST invariant  restrictions 
on the anti-chiral superfields (cf. Sec 3). 
The substitution  of the above expressions  for the secondary fields (cf. (24))
into the super expansions\footnote { We have taken the coefficient of $\theta$, in the super expansion of the appropriate superfields,  as
the anti-BRST symmetry transformation on a suitable {\it ordinary} field because this has been explicitly and precisely
proven in the earlier works [1-15].} (21) yields the following expressions
\begin{eqnarray}
B_{\mu}^{(ab)} (x, \theta) & = & A_{\mu}(x) + \theta\; (\partial_\mu\bar C)\;\quad \equiv~~ \; A_{\mu}(x) + \theta\;(s_{ab}\; A_\mu(x)),\nonumber\\
F^{(ab)}(x, \theta) &=& C(x) + \theta\;(-\;i\;B)\quad\equiv~~ C(x) + \theta\ (s_{ab}\; C(x)),\nonumber\\
\bar F^{(ab)}(x, \theta) &=& \bar C(x) + \theta\;(0)\qquad\quad\,\equiv ~~ \bar C(x) + \theta\ (s_{ab} \;\bar C(x)),\nonumber\\\
\Psi ^{(ab)}(x, \theta) &=& \psi (x) + \theta\;(-\,i\,e\,\bar C\,\psi )\equiv ~~ \psi (x) + \theta\ (s_{ab}\; \psi(x)),\nonumber\\
\bar\Psi ^{(ab)}(x, \theta) &=& \bar\psi (x) + \theta\;(-\,i\,e\,\bar\psi \,\bar C)\equiv ~~ \bar\psi (x) + \theta\ (s_{ab}\; \bar\psi(x)),\nonumber\\
\tilde B^{(ab)} (x, \theta) &=& B(x) + \,\theta\,(0) \qquad \quad\equiv ~~  B(x) + \theta\;(s_{ab}\; B (x)),
\end{eqnarray}
where the superscript $(ab)$ denotes the super expansions of the chiral superfields after the 
application of anti-BRST invariant restrictions in Eq. (23).

We would like to end this section with the following remarks.
First, we have derived the anti-BRST symmetry transformations 
of (2) by invoking the anti-BRST invariant restrictions 
on the superfields (cf. Eqs. (22), (23)). Second, the anti-BRST 
symmetry of an {\it ordinary} field corresponds to the translation 
of corresponding {\it chiral} superfield  
(obtained after the application of anti-BRST invariant restrictions)
along  $\theta$-direction
of the (D, 1)-dimensional {\it chiral} super-submanifold (of the 
general (D, 2)-dimensional  supermanifold).
Finally, the nilpotency ($s_{ab}^2 = 0$) of the 
anti-BRST  symmetry transformation is intimately connected with
the nilpotency ($\partial_{\theta}^2 = 0$) of the 
translation generators ($\partial_{\theta}$)
along $\theta$-direction of the {\it chiral} super-submanifold.
Similar kinds of observations can be made and stated for the 
BRST symmetries, too, in the language of  ACSA to BRST formalism
(cf. Sec. 3).

\section 
{\bf Conserved (Anti-)BRST Charges: Nilpotency and Absolute Anicommutativity Properties}

In this section, we shall capture the properties of nilpotency and
absolute anticommutativity of the conserved (anti-)BRST charges
in the language of  ACSA to BRST formalism. It is straightforward to express the 
BRST charge $\big(Q_b =  \int\; d^{D-1} x\; \big[ B\;\dot C - \dot B \; C \big] \big)$
in terms of the {\it anti-chiral} superfields and partial derivative $\partial_{\bar\theta}$ and/or differential $d\bar\theta$ as
\begin{eqnarray}
Q_b  & = &\frac {\partial}{\partial\bar\theta}\;\int\;d^{D-1}x\;\Big[i\;\dot{\bar F}^{(b)}(x,\bar\theta) F^{(b)}(x,\bar\theta)
- i\; \bar F^{(b)}(x,\bar\theta)\dot F^{(b)}(x,\bar\theta)\Big]\nonumber\\
&\equiv & \int\; d \bar\theta\int\;d^{D-1}x\;\Big[i\;\dot{\bar F}^{(b)}(x,\bar\theta) F^{(b)}(x,\bar\theta)
- i\; \bar F^{(b)}(x,\bar\theta)\dot F^{(b)}(x,\bar\theta)\Big],
\end{eqnarray}
where the superscript $(b)$ stands for the anti-chiral superfields that have been obtained 
after the application of the BRST invariant restrictions (11).
Thus, the nilpotency ($\partial_{\bar\theta}^ 2 = 0$) of the translational generator
$\partial_{\bar\theta}$ implies that we have\footnote{From the super expansions
(20) and (25), it is evident that we have established, in our
present endeavor, a relationship
between the (anti-)BRST  symmetry transformations and the translational generators 
$(\partial_{\theta}, \partial_{\bar\theta})$ along the Grassmannian directions
$(\theta,\bar\theta)$ of the chiral and anti-chiral (D, 1)-dimensional
super-submanifolds (of the general (D, 2)-dimensional supermanifold), respectively.} 
\begin{eqnarray}
&&\partial_{\bar\theta}\; Q_b  = 0\quad\Longleftrightarrow \quad \partial_{\bar\theta}^ 2 = 0\quad\Longleftrightarrow \quad s_b\; Q_b  = -\;i\;
{\{Q_b, Q_b}\} = 0,
\end{eqnarray}
which implies the nilpotency ($Q_b^2 = 0$)
of the BRST charge $Q_b$. Thus, we have shown that there is a deep connection 
between the nilpotency ($\partial_{\bar\theta}^ 2 = 0$) of the translational
generator $(\partial_{\bar\theta})$ and the nilpotency (i.e. 
$Q_b^2 = 0$) of the BRST charge ($Q_b$).

Now we dwell a bit on the absolute anticommutativity property of 
the BRST charge ($Q_b$) with the anti-BRST charge ($Q_{ab}$). It can be readily checked that the BRST charge 
$Q_b$ can {\it also} be expressed in terms of the {\it chiral} superfields
as 
\begin{eqnarray}
 Q_b  & = & \frac {\partial}{\partial\theta}\int\; d^{D-1}x\;\Big[i\;F^{(ab)}(x,\theta)\dot F^{(ab)}(x,\theta)\Big]\nonumber\\
& \equiv & \int\; d\theta\;\int\; d^{D-1}x\;\Big[i\;F^{(ab)}(x,\theta)\dot F^{(ab)}(x,\theta)\Big],
\end{eqnarray}
where the superscript $(ab)$ denotes the {\it chiral} superfields that have been obtained after the application of the 
anti-BRST invariant restrictions in Eq. (23).
It is now straightforward to check that we have the following:
\begin{eqnarray}
&&\partial_{\theta}\; Q_b  = 0\quad\Longleftrightarrow \quad \partial_{\theta}^ 2 = 0\quad\Longleftrightarrow \quad s_{ab}\; Q_b  = -\;i\;
{\{Q_b, Q_{ab}}\} = 0,
\end{eqnarray}
which proves the absolute anticommutativity of the BRST charge with anti-BRST charge.
It is interesting to point out that the above anticommutativity is connected with the 
nilpotency of the translational generator $(\partial_{\theta})$. Thus,
we observe that, for the BRST charge $(Q_b)$, the nilpotency ($ Q_b^2 = 0$) is connected with the 
nilpotency $(\partial_{\bar\theta}^2 = 0)$  of the translational  generator
($\partial_{\bar\theta}$) along the 
{\it anti-chiral}  super-submanifold {\it but} the absolute anticommutativity of the 
BRST charge $(Q_b)$ with the anti-BRST  charge $(Q_{ab})$ is intimately 
related to the nilpotency ($\partial_{\theta}^2 = 0$)  of the translational generator $(\partial_{\theta})$ along 
the $\theta$-direction of the {\it chiral} (D, 1)-dimensional super-submanifold. These are completely  {\it novel} and interesting 
observations.

We concentrate now on the off-shell nilpotency and absolute anticommutativity
of the anti-BRST charge. The nilpotency can be expressed in the language of the 
chiral superfields  and the Grassmannian partial derivative $(\partial_{\theta})$ as well
as the differential ($d\theta$) as:
\begin{eqnarray}
&& Q_{ab} = \frac {\partial}{\partial\theta}\int\; d^{D-1}x\;\Big[ i\;\bar F^{(ab)}(x,\theta)\;\dot F^{(ab)}(x,\theta)  
- i\;\dot {\bar F}^{(ab)}(x,\theta)\;F^{(ab)}(x,\theta)\Big]\nonumber\\
&&\equiv \int\; d\theta\;\int\; d^{D-1}x\;\Big[ i\;\bar F^{(ab)}(x,\theta)\;\dot F^{(ab)}(x,\theta)  
- i\;\dot {\bar F}^{(ab)}(x,\theta)\;F^{(ab)}(x,\theta)\Big],
\end{eqnarray}
where the chiral superfields with superscript $(ab)$ have been expressed in Eq. (25).
It is straightforward to note that the following connections are true, namely;
\begin{eqnarray}
&&\partial_{\theta}\;Q_{ab} = 0\quad\Longleftrightarrow \quad \partial_{\theta}^2 = 0\quad\Longleftrightarrow \quad s_{ab}\;Q_{ab}
 = - i\;{\{Q_{ab}, Q_{ab}}\} = 0.
\end{eqnarray}
The last entry, in the above equation, implies the nilpotency ($Q_{ab}^2 = 0$)
of the anti-BRST charge $Q_{ab}$. It is also evident that the nilpotency
of the translational generators ($\partial_\theta$) is deeply connected 
with the nilpotency of the anti-BRST charge. The absolute anticommutativity
of the anti-BRST charge $(Q_{ab})$ with the BRST charge $(Q_b)$ can be written 
as 
\begin{eqnarray}
 Q_{ab} & = &\frac {\partial}{\partial\bar\theta}\;\int\;d^{D-1}x\;[- \; \bar F^{(b)}(x,\bar\theta)\; \dot {\bar F}^{(b)}(x,\bar\theta)
\Big]\nonumber\\
&\equiv & \int\; d\bar\theta\;\int\;d^{D-1}x\;[- \; \bar F^{(b)}(x,\bar\theta)\; \dot {\bar F}^{(b)}(x,\bar\theta)\Big],
\end{eqnarray}
where the anti-chiral superfields with superscript $(b)$ are written in (20).
It is obvious that the following mapping is  true, namely; 
\begin {eqnarray}
&&\partial_{\bar\theta}\;Q_{ab} = 0\quad\Longleftrightarrow \quad \partial_{\bar\theta}^2 = 0\quad\Longleftrightarrow \quad s_b\;Q_{ab}
 = - i\;{\{Q_{ab}, Q_b}\} = 0,
\end{eqnarray}
which establishes the absolute anticommutativity ${\{Q_{ab}, Q_b}\} = 0$ of the (anti-)BRST
charges $Q_{(a)b}$. From our discussions, it is clear that the nilpotency ($Q_{ab}^2 = 0$) of the 
anti-BRST charge  ($Q_{ab}$) is connected with the nilpotency ($\partial_{\theta}^2 = 0$)
of the translational generator $\partial_{\theta}$ along  $\theta$-direction
of the {\it chiral} super-submanifold {\it but } the absolute anticommutativity 
of the anti-BRST charge with the  BRST charge is intimately related with the
nilpotency  ($\partial_{\bar\theta}^2 = 0$) of the translational 
generator $\partial_{\bar\theta}$ along  ${\bar\theta}$-direction of 
the {\it anti-chiral} super-submanifold of the (D, 2)-dimensional supermanifold
on which our ordinary theory is generalized.

The above properties of the nilpotency and absolute anticommutativity
of the (anti-) BRST charges $Q_{(a)b}$,  discussed within the 
framework of ACSA to BRST formalism, can be expressed in the {\it ordinary}
space in terms of the (anti-)BRST symmetry transformations $s_{(a)b}$
(due to their connection with $\partial_\theta$ and $\partial_{\bar\theta}$).
In other words, these aspects (i.e. nilpotency and absolute anticommutativity)  
of the conserved (anti-)BRST charges can be easily proven
 due to the following (anti-)BRST {\it exact} forms of them, namely;
\begin{eqnarray}
&&Q_b = s_b \int d^{D-1}x\;[i\,\dot{\bar C}\,C - i\,\bar C\,\dot C], \qquad\quad Q_b = s_{ab} \int d^{D-1}x\;(i\,C\,\dot C),\nonumber\\
&&Q_{ab} = s_{ab} \int d^{D-1}x\;[i\,\bar C\,\dot C - i\,\dot{\bar C}\, C],\qquad Q_{ab} = s_b \int d^{D-1}x\;(-\,i\,\bar C\,\dot{\bar C}).
\end{eqnarray}
Applying the symmetry principle on the fermionic operators (cf. Eq. (8)),
we obtain:
\begin{eqnarray}
&&s_b Q_b = -i\,{\{Q_b,Q_b}\} = 0\qquad\quad \longleftrightarrow  \qquad\qquad\; Q_b^2 = 0\quad \Longleftrightarrow\quad s{_b}^2 = 0,\nonumber\\
&&s_{ab} \,Q_{ab} = -i\,\{Q_{ab},Q_{ab}\} = 0 \quad\; \longleftrightarrow \qquad\qquad Q_{ab}^2 = 0\quad \Longleftrightarrow \quad s_{ab}^2 = 0,\nonumber\\
&&s_b Q_{ab}  = -i\,{\{Q_{ab},Q_b}\} = 0 \quad\quad\;\longleftrightarrow   \;\;{\{Q_{ab},Q_b}\} = 0 \quad\;\;\; \Longleftrightarrow  \quad s_{b}^2 = 0,\nonumber\\
&&s_{ab}Q_b = -i\,{\{Q_b,Q_{ab}}\} = 0\quad\quad \,\longleftrightarrow    \;\, {\{Q_b,Q_{ab}}\} = 0\qquad \Longleftrightarrow \quad  s_{ab}^2 = 0.
\end{eqnarray}
Thus, we note that, because of the (anti-)BRST {\it exact} forms in (34), we are able to prove the off-shell nilpotency 
as well as absolute anticommutativity of the conserved (anti-)BRST charges $Q_{(a)b}$. The key and crucial role, in the above 
proof, is played by the concept behind the continuous symmetries and their generators (as the conserved Noether charges).
This idea has been backed and bolstered by the nilpotency ($s_{(a)b}^2 = 0$)  of the (anti-)BRST symmetry transformations as is evident from Eq. (35).
We would like to lay {\it emphasis} on the fact that Eqs. (34) and (35) have been derived due to 
our knowledge of ACSA to BRST formalism because the 
key equations (26), (28), (30) and (32) are responsible for their
derivations.

\section
{\bf (Anti-)BRST Invariance: Superfield Approach}

We have seen that the Lagrangian density (1) of our Sec. 2 respects the 
infinitesimal, continuous and nilpotent (anti-)BRST symmetries because
this Lagrangian density transforms to the total spacetime derivatives 
under the above symmetries. As a consequence, the action integral (corresponding to
this Lagrangian density) respects the (anti-)BRST symmetries in a precise and perfect manner.
In our present section,  
we briefly capture  the (anti-)BRST invariance  (cf. Eq. (3))
of the Lagrangian density (1) in the language of ACSA to BRST 
formalism\footnote{In other words, first of all, we express the Lagrangian density (1) in the 
language of the (anti-) chiral  superfields (that are defined on the 
(anti-)chiral super-submanifolds) and study their key properties by applying the translational generators 
$(\partial_\theta, \partial_{\bar\theta})$ on them.}. In this context, we note that the ordinary Lagrangian density 
${\cal L}_B$ 
can be generalized onto the (anti-)chiral super-submanifolds in terms of the
(anti-)chiral superfields  as 
\begin{eqnarray}
{\cal L}_B\longrightarrow \tilde {\cal L}^{(ac)}_B & = & -\;\frac {1}{4}\; \tilde F^{\mu\nu( b)}(x, \bar\theta)\;\tilde F_{\mu\nu}^{( b)}(x, \bar\theta)
+ \bar\Psi^{(b)}(x, \bar\theta)\;( i\;\gamma^{\mu}\,\partial_{\mu} - m)\;\Psi^{(b)}(x, \bar\theta)\nonumber\\
& - & e\;\bar\Psi^{(b)}(x, \bar\theta)\;\gamma^{\mu}\; B_{\mu}^{(b)}(x, \bar\theta)\;\Psi^{(b)}(x, \bar\theta) + \tilde B^{(b)}(x, \bar\theta)
(\partial_{\mu} B^{\mu (b)}(x, \bar\theta))\nonumber\\
& + & \frac {1}{2}\; \tilde B^{(b)}(x, \bar\theta) \;\tilde B^{(b)}(x, \bar\theta) -\; i\; \partial_{\mu}\bar F^{(b)}(x, \bar\theta)\;\partial^{\mu} F^{(b)}(x, \bar\theta),\nonumber\\
{\cal L}_B\longrightarrow \tilde {\cal L}^{(c)}_B & = & -\;\frac {1}{4}\; \tilde F^{\mu\nu (ab)}(x, \theta)\;\tilde F_{\mu\nu}^{(a b)}(x, 
\theta)
+ \bar\Psi^{(ab)}(x, \theta)\;( i\;\gamma^{\mu}\,\partial_{\mu} - m)\;\Psi^{(ab)}(x, \theta)\nonumber\\
& - & e\;\bar\Psi^{(ab)}(x, \theta)\;\gamma^{\mu}\; B_{\mu}^{(ab)}(x, \theta)\;\Psi^{(ab)}(x, \theta) + \tilde B^{(ab)}(x, \theta)
(\partial_{\mu} B^{\mu (ab)}(x, \theta))\nonumber\\
& + & \frac {1}{2}\; \tilde B^{(ab)}(x, \theta) \;\tilde B^{(ab)}(x, \theta) -\;i \; \partial_{\mu}\bar F^{(ab)}(x, \theta)\;\partial^{\mu} F^{(ab)}(x, \theta),
\end{eqnarray}
where $\tilde B^{(ab)}(x, \theta) = B(x)$ $\equiv \tilde B^{(b)}(x, \bar\theta)$, $F^{(b)}(x, \bar\theta) = C(x)$,
$\bar F^{(ab)}(x, \theta) = \bar C(x)$, $\tilde F^{\mu\nu(b)}(x, \bar\theta) = F_{\mu\nu}(x) = \tilde F^{\mu\nu(ab)}(x, \theta)$
because of the fact that {\it all} these quantities are (anti-)BRST
invariant (i.e. $s_{(a)b} B = 0,\;  s_b C = 0,\; s_{ab}\bar C = 0,\;s_{(a)b} F_{\mu\nu} = 0$).
Thus, the above anti-chiral super Lagrangian density $\tilde{\cal L}_B^{(ac)}$ and chiral super Lagrangian density $\tilde{\cal L}_B^{(c)}$
can be re-written as:
\begin{eqnarray}
\tilde {\cal L}^{(ac)}_B & = & -\;\frac {1}{4}\;  F^{\mu\nu}(x)\; F_{\mu\nu}(x)
+ \bar\Psi^{(b)}(x, \bar\theta)\;( i\;\gamma^{\mu}\,\partial_{\mu} - m)\;\Psi^{(b)}(x, \bar\theta)\nonumber\\
& - & e\;\bar\Psi^{(b)}(x, \bar\theta)\;\gamma^{\mu}\; B_{\mu}^{(b)}(x, \bar\theta)\;\Psi^{(b)}(x, \bar\theta) +  B(x)
(\partial_{\mu} B^{\mu (b)}(x, \bar\theta))\nonumber\\
& + & \frac {1}{2}\;  B^2(x) -\; i\; \partial_{\mu}\bar F^{(b)}(x, \bar\theta)\;\partial^{\mu} C(x),
\end{eqnarray}
\begin{eqnarray}
\tilde {\cal L}^{(c)}_B & = & -\;\frac {1}{4}\;  F^{\mu\nu}(x)\; F_{\mu\nu}(x)
+ \bar\Psi^{(ab)}(x, \theta)\;( i\;\gamma^{\mu}\,\partial_{\mu} - m)\;\Psi^{(ab)}(x, \theta)\nonumber\\
& - & e\;\bar\Psi^{(b)}(x, \theta)\;\gamma^{\mu}\; B_{\mu}^{(ab)}(x, \theta)\;\Psi^{(ab)}(x, \theta) +  B(x)
(\partial_{\mu} B^{\mu (ab)}(x, \theta))\nonumber\\
& + & \frac {1}{2}\;  B^2 (x) -\;i \; \partial_{\mu}\bar C(x)\;\partial^{\mu}\bar F^{(ab)}(x, \theta).
\end{eqnarray}
It is worthwhile to mention here that the following are true, namely; 
\begin{eqnarray}
&&  \frac   {\partial}{\partial\bar\theta}\; \Big [  \bar\Psi^{(b)}(x, \bar\theta)\;( i\;\gamma^{\mu}\,\partial_{\mu} - m)\;\Psi^{(b)}(x, \bar\theta)
 - \; e\;\bar\Psi^{(b)}(x, \bar\theta)\;\gamma^{\mu}\; B_{\mu}^{(b)}(x, \bar\theta)\;\Psi^{(b)}(x, \bar\theta)\Big] = 0\nonumber\\
&& \Longrightarrow s_b \;\Big [  \bar\psi\;( i\;\gamma^{\mu}\,D_{\mu} - m)\;\psi\Big] = 0\quad \mbox {and} \quad 
\frac   {\partial}{\partial\theta}\; \Big [ \bar\Psi^{(ab)}(x, \theta)\;( i\;\gamma^{\mu}\,\partial_{\mu} - m)\;\Psi^{(ab)}(x, \theta)\nonumber\\
 &&- \; e\;\bar\Psi^{(ab)}(x, \theta)\;\gamma^{\mu}\; B_{\mu}^{(ab)}(x, \theta)\;\Psi^{(ab)}(x, \theta)\Big] = 0\nonumber\\
 &&\Longrightarrow s_{ab} \;\Big [  \bar\psi\;( i\;\gamma^{\mu}\,D_{\mu} - m)\;\psi\Big] = 0,
\end{eqnarray}
due to the fact that $ s_{(a)b} \big[   \bar\psi\;( i\;\gamma^{\mu}\,D_{\mu} - m)\;\psi\big] = 0$.

Now we are in the position to state the (anti-)BRST invariance (cf. Eq. (3))
of the Lagrangian density (1) in the language of ACSA to BRST formalism.
Taking the inputs from (39), we derive the following interesting results, namely;
\begin{eqnarray}
&& \frac {\partial}{\partial\theta}\Big[\tilde {\cal L}_B^{(c)}\Big] = \partial_{\mu}\big(B\;\partial^{\mu}\bar C\big),\qquad
\frac {\partial}{\partial\bar\theta}\Big[\tilde {\cal L}_B^{(ac)}\Big] = \partial_{\mu}\big(B\;\partial^{\mu} C\big),
\end{eqnarray}
which establish the geometrical interpretation of the 
(anti-)BRST invariance, quoted in Eq. (3), in the following manner.
The translation of the super Lagrangian density $\tilde {\cal L}_B^{(ac)}$ along $\bar\theta$-direction
of the (D, 1)-dimensional {\it anti-chiral} super-submanifold produces the total spacetime derivative in the
{\it ordinary} space. Similarly, the translation of the super Lagrangian density  $\tilde {\cal L}_B^{(c)}$
along $\theta$-direction of the (D, 1)-dimensional {\it chiral} super-submanifold leads to the 
derivation of a total spacetime derivative term (cf. (40)) 
thereby rendering the action integral 
invariant in the {\it ordinary} space for our present {\it interacting} Abelian theory with 
Dirac's fields $(\psi, \bar\psi)$.

\section 
{\bf Conclusions}

In our present investigation, we have discussed  the nilpotency and absolute anticommutativity 
properties of the conserved (anti-)BRST charges of the {\it ordinary}  D-dimensional {\it interacting} Abelian
1-form gauge theory (where there is a coupling between the gauge field and the Dirac fields) in the language 
of ACSA to BRST formalism. The {\it novel} observation of our present endeavor is the proof of the 
{\it absolute anticommutativity} of the (anti-)BRST charges  {\it despite } the fact that
we have taken into account {\it only} the (anti-)chiral super expansions of the supefields (cf. Sec. 5).
We have shown that these observations/results are {\it also} true when there is a coupling between the $U(1)$
 gauge field and the complex scalar fields (cf. Appendix A).
The nilpotency and absolute anticommutativity of the above charges and 
corresponding continuous (anti-)BRST symmetries are {\it obvious } when the {\it full} super
expansion of the superfields is taken into account (cf.  Appendix B).

It is interesting to note that the nilpotency of the BRST and anti-BRST 
charges is connected with the nilpotency of the translational generators 
$\partial_{\bar\theta}$ and $\partial_\theta$, respectively. However, 
we have established (cf. Sec. 5) that the absolute anticommutativity
of the BRST charge with anti-BRST charge is connected with the nilpotency
($\partial_{\theta}^2 = 0$) of the translational generator ($\partial_\theta$)
along the $\theta$-direction of the {\it chiral} super-submanifold. On the contrary, the absolute anticommutativity
of the anti-BRST charge with BRST charge is connected with the nilpotency 
($\partial_{\bar\theta}^2 = 0$) of the translational generator ($\partial_{\bar\theta}$) along $\bar\theta$-direction
of the {\it anti-chiral} super-submanifold. These observations are completely {\it novel} as far as various forms of
superfield approaches to BRST formalism are concerned (see, e.g. [1-15]). These observations 
can be stated in the language of the nilpotency $(s_{(a)b}^2 = 0)$ of the (anti-)BRST
symmetry transformations $(s_{(a)b})$ in a straightforward fashion (cf. Sec. 5).

We envisage the extension of our present idea in the context of 
D-dimensional  non-Abelian 1-form gauge theory where there would be
{\it interaction} between $SU(N)$ non-Abelian gauge field and the matter 
fields (i.e. Dirac fields) in any arbitrary dimension of spacetime [20] where the celebrated 
Curci-Ferrari condition [23] would play very important role. 
Furthermore, for the 2D non-Abelian gauge theory, we have shown the 
existence of nilpotent (anti-)co-BRST charges (see, e.g. [24]) in addition to the (anti-)BRST
charges. We would like to capture the nilpotency and absolute anticommutativity
of the  (anti-)co-BRST charges within the framework of ACSA to BRST formalism
as we have done for the (anti-) BRST charges in the case of our present { \it interacting}  Abelian 1-form theory
with Dirac's fields $(\psi, \bar\psi)$.
These are the issues that would be discussed in our future investigations [25].\\

\noindent
{\bf Acknowledgments:}
B. Chauhan and S. Kumar are grateful to the DST-INSPIRE and BHU fellowships for financial support
under which the present work has been carried out.\\

\begin{center}
{\bf Appendix A: Interacting Abelian 1-Form  Theory with Complex Scalar Fields: ACSA to BRST formalism}\\
\end{center}

\noindent
Here we discuss about the D-dimensional {\it interacting} Abelian 1-form gauge 
theory where there is a coupling between the $U(1)$ gauge field $(A_\mu)$ and the 
complex scalar fields $(\phi, \phi^\ast)$. Our objective is to exploit the 
beauty and strength of the (anti-)chiral superfield approach to derive the (anti-)BRST
symmetry transformations (for this {\it interacting} gauge theory). Towards this 
goal in mind, we begin with the dynamically closed system of $(A_\mu)$ and $(\phi, \phi^\ast)$
fields which is described by the following (anti-)BRST invariant Lagrangian density  (with $F_{\mu\nu} =
\partial_\mu A_\nu-\partial_\nu A_\mu $)
\[{\cal L}_b = - \frac {1}{4}F^{\mu\nu}F_{\mu\nu} + (D_\mu\phi)^\ast (D^\mu\phi)- m^2\phi^\ast\phi + B\;(\partial\cdot A) + \frac {1}{2}B^2 - i\;\partial_\mu{\bar C}\; \partial^\mu C,\eqno(A.1)\]
where the covariant derivatives: $(D_\mu\phi)^\ast\equiv \bar D_\mu\phi^\ast =(\partial_\mu- i\,e A_\mu)\,\phi^\ast$ 
and $D_\mu\phi = (\partial_\mu + i\,e A_\mu)\,\phi$
are defined for fields $( \phi^\ast, \phi)$. Here $m$ is the rest mass of
the complex scalar fields and B is the Nakanishi Lautrup auxiliary field and 
$(\bar C )C$ are the fermionic $(C^2 = \bar C^2 = 0, C\bar C+\bar C C  = 0)$ (anti-)ghost fields. It is elementary to check that the following 
infinitesimal, continuous, nilpotent ($s_{(a)b}^2 = 0$)
and absolutely anticommuting $(s_b s_{ab} + s_{ab} s_b = 0)$ (anti-) BRST transformations $(s_{(a)b})$
\[s_{ab} A_\mu = \partial_\mu \bar C,\qquad\;\; s_{ab} \bar C = 0,\qquad\qquad\;\; s_{ab} C = -\,i\,B,\qquad s_{ab} B = 0,\]
\[s_{ab}\phi = -\,i\,e\,\bar C\phi,\quad\; s_{ab} \phi^\ast  = +\,i\,e\,\phi^\ast\;\bar C ,\quad\, s_{ab }F_{\mu\nu} = 0,\quad s_{ab}(\partial\cdot A) = \square\bar C,\]
\[s_b A_\mu = \partial_\mu C,\qquad\quad s_b C = 0,\qquad\qquad\;\;\; s_b \bar C = i\,B,\qquad\quad\;\,  s_b B = 0,\]
\[s_b \phi = -\,i\,e\,\, C\phi,\quad\;\; s_b \phi^\ast  = +\,i\,e\,\phi^\ast C\, \quad\;\;\, s_b F_{\mu\nu} = 0,\quad s_b (\partial\cdot A) = \square C,\eqno(A.2)\]
leave the action integral $S = \int d^{D}x \;{\cal L}_b $ invariant because the Lagrangian density $({\cal L}_b)$
transforms, under the above nilpotent (anti-)BRST symmetry transformations, {\it exactly} as given in Eq. (3) (i.e. $s_b\;{\cal L}_b  = \partial_\mu (B\,\partial^\mu C),\;s_{ab}\;{\cal L}_ b  = \partial_\mu (B\,\partial^\mu \bar C)$).

According to celebrated Noether's theorem, we have the following expressions for the 
conserved currents:
\[J^{\mu}_{ab}  = - F^{\mu\nu}\partial_{\nu}\bar C + B\; \partial^{\mu}\bar C - i\; e\;\bar C\, \phi \; (D^{\mu}\phi)^\ast + i\, e\,\bar C \phi ^\ast D^\mu \phi,\]
\[J^{\mu}_b  = - F^{\mu\nu}\partial_{\nu} C + B\; \partial^{\mu} C - i\; e\; C \phi \; (D^{\mu}\phi)^\ast + i\, e\, C\, \phi ^\ast D^\mu \phi.\eqno(A.3)\]
The conservation law $\partial_\mu J^\mu_{(a)b} = 0$ can be proven due to the following EL-EOMs that are
derived from Lagrangian density ${\cal L}_b$:
\[\partial_{\mu}F^{\mu\nu} - \partial^\nu B - i\; e\; \phi^\ast D^{\nu}\phi+ i\; e\; (D^{\nu}\phi)^\ast\phi = 0,\quad
 \Box C = 0 = \Box\bar C,\]
\[\bar D_\mu (D^\mu\phi)^\ast = -\; m^2\phi^\ast,\quad D_\mu (D^\mu\phi) =  -\; m^2\phi,\qquad B = - (\partial\cdot A).\eqno(A.4)\]
According to, once again, the Noether theorem, one can define  the conserved charges corresponding to
the (anti-)BRST currents (cf. Eq. (A.3)) as:
\[Q_{ab} = \int d^{D-1}x\,J_{ab}^0 = \int d^{D-1}x\; [ - F^{0i}\partial_i \bar C + B\;\dot{\bar C} - i\,e\,\bar C\,\phi (D^0\phi)^\ast + i\;e\;\bar C\;\phi^\ast D^0\phi],\]
\[Q_b = \int d^{D-1}x\,J_b^0 = \int d^{D-1}x \;[ - F^{0i}\partial_i C + B\;\dot C - i\,e\, C\,\phi (D^0\phi)^\ast + i\;e\; C\;\phi^\ast D^0\phi].\eqno(A.5)\]
Using the EL-EOMs from (A.4) (i.e. $\partial_i F^{0i} = - (\dot B + i\;e\;\phi^\ast D^0\phi - i\;e\;(D^0\phi)^\ast\phi))$
and exploiting the beauty of Gauss's divergence theorem, we obtain the expressions for the conserved (anti-)BRST
charges $(Q_{(a)b})$ which are exactly  {\it same} as the {\it final} expressions for them  in Eqs. (6) and (7) (i.e. $Q_{ab} = \int d^{D-1}x \,( B\;\dot{\bar C} - \dot B \; \bar C)$ and $Q_b = \int d^{D-1}x \,( B\;\dot C - \dot B \;  C))$.

To derive the (anti-)BRST symmetry transformations (A.2), we exploit the (anti-)chiral
super expansions (9) and (21) except that {\it now} the matter fields $(\phi, \phi^\ast)$ would have the 
following generalizations and super expansions
\[\phi (x)\longrightarrow \Phi (x, \bar\theta) = \phi (x) + i\,\bar\theta\, P_1 (x), \qquad
\phi (x)\longrightarrow \Phi (x, \theta) = \phi (x) + i\,\theta \,\bar P_1 (x),\]
\[\phi ^\ast (x)\longrightarrow \Phi ^\ast (x, \bar\theta) = \phi^\ast (x) + i\,\bar\theta\; P_2 (x),
 \qquad\phi ^\ast (x)\longrightarrow \Phi ^\ast (x, \theta) = \phi^\ast (x) + i\,\theta\; \bar P_2 (x),\eqno(A.6)\]
 where ($P_1 (x),P_2 (x), \bar P_1 (x), \bar P_2 (x)$) are the secondary fields which are to be determined by the (anti-)BRST invariant restrictions.
 In this context, we note that the following are the {\it useful} and interesting  (anti-)BRST invariant quantities:
\[s_{ab} B = 0,\qquad s_{ab} (\bar C\;\phi) = 0, \qquad s_{ab} (\phi^ \ast \bar C) = 0,\qquad s_{ab} (\phi ^\ast \phi ) = 0,\]
\[s_{ab} (\phi ^\ast D_\mu \phi) = 0,\qquad s_{ab} (\phi  \bar D_\mu \phi ^\ast) = 0,\qquad s_{ab} \bar C = 0,\qquad s_{ab}(\bar D_\mu\phi^\ast D^\mu\phi)= 0,\]
\[s_{ab} (A^\mu\partial_\mu \bar C) = 0, \qquad s_{ab} [A^\mu \partial_\mu B + i\,\partial_\mu \bar C \partial ^\mu C] = 0,\]
\[s_b B = 0,\qquad s_b (C\,\phi) = 0, \qquad s_b (\phi^ \ast C) = 0,\qquad s_b (\phi ^\ast\;\phi) = 0,\]
\[s_b (\phi ^\ast D_\mu \phi) = 0,\quad s_b (\phi  \bar D_\mu \phi ^\ast) = 0,\quad s_b C = 0,\qquad s_b(\bar D_\mu\phi^\ast D^\mu\phi)= 0,\]
\[s_b (A^\mu\partial_\mu C) = 0, \qquad s_b [A^\mu \partial_\mu B + i\,\partial_\mu \bar C \partial ^\mu C] = 0.\eqno(A.7)\]
 Using the basic tenets of augmented (anti-)chiral superfield approach, we shall obtain the super 
expansions of (anti-)chiral superfields corresponding to the auxiliary, gauge and fermionic (anti-)ghost fields as given in Eqs. 
(20) and (25). We shall exploit these expansions to determine the secondary fields ($P_1 (x),P_2 (x), \bar P_1 (x), \bar P_2 (x)$)
so that we could obtain the (anti-)BRST symmetry transformations for the matter fields (i.e. complex scalar fields).
Towards  this objective in mind, we observe that the following are true (if we take into account the expansions of Eqs. (20) and (25)): 
\[s_b (C\,\phi) = 0\Longrightarrow F^{(b)} (x,\bar\theta)\; \Phi (x,\bar\theta) = C(x)\,\phi (x)\Longrightarrow 
 C\;P_1   = 0, \]
\[s_b (\phi^\ast C) = 0 \Longrightarrow \Phi^\ast (x,\bar\theta)\; F^{(b)}(x,\bar\theta) = \phi^\ast (x)\; C(x)\Longrightarrow 
P_2\; C  = 0,\] 
\[s_{ab} (\bar C\,\phi) = 0\Longrightarrow \bar F^{(ab)} (x,\theta) \;\Phi (x,\theta) = \bar C(x)\;\phi (x)\Longrightarrow 
\bar C\;\bar P_1   = 0, \]  
\[s_{ab} (\phi^\ast \bar C) = 0 \Longrightarrow \Phi^\ast (x,\theta)\; \bar F^{(ab)}(x,\theta) = \phi^\ast (x)\; \bar C(x)\Longrightarrow 
\bar P_2 \;\bar C  = 0.\eqno(A.8)\]
The above restrictions imply that the non-trivial solutions are: $P_1 \propto C$, $P_2 \propto C$, $\bar P_1 \propto \bar C$, $\bar P_2 \propto \bar C$.
To obtain the explicit forms of ($P_1, P_2, \bar P_1,\bar P_2$), we have to exploit some of the  other key restrictions of (A.7).
For instance, we shall exploit now $s_{(a)b} (\phi ^\ast D_\mu \phi) = 0$, $s_{(a)b} (\phi  \bar D_\mu \phi ^\ast) = 0$
which imply the following restrictions

\[\Phi^\ast (x, \bar\theta)\;\partial_\mu \Phi (x, \bar\theta) + i\,e\,\Phi^\ast (x, \bar\theta)\; B_{\mu}^{(b)}(x, \bar\theta)\;\Phi (x, \bar\theta) = 
\phi^\ast (x)\; D_\mu \phi (x), \]
\[\Phi (x, \bar\theta)\;\partial_\mu \Phi^\ast (x, \bar\theta)- i\,e\,\Phi (x, \bar\theta) B_{\mu}^{(b)}(x, \bar\theta)\;\Phi^\ast (x,\bar\theta)=
\phi (x)\; \bar D_\mu \phi^\ast (x),\]
\[\Phi^\ast (x, \theta)\;\partial_\mu \Phi (x, \theta) + i\,e\,\Phi^\ast (x, \theta)\; B_{\mu}^{(ab)}(x, \theta)\;\Phi (x, \theta)  = 
\phi^\ast (x)\; D_\mu \phi (x),\] 
\[\Phi (x, \theta)\;\partial_\mu \Phi^\ast (x, \theta) - i\,e\,\Phi (x, \theta)\; B_{\mu}^{(ab)}(x, \theta)\;\Phi^\ast (x, \theta)  = 
\phi (x)\; \bar D_\mu \phi^\ast (x).\eqno(A.9)\] 
Taking the helps from (20) and (25), where the explicit forms of $B_{\mu}^{(b)}(x,\bar\theta)$ and  $B_{\mu}^{(ab)}(x,\theta)$ are given,
we obtain the exact expressions for the secondary fields in terms of the (anti-)ghost fields and complex scalar fields
of our present {\it interacting} Abelian 1-form gauge theory.

To corroborate the above statement, we explicitly compute the first two lines of restrictions that are 
quoted in Eq. (A.9). We obtain the following relationships from these restrictions
(if we take the helps from Eq. (A.6) and Eq. (20)):
\[i\;\phi^\ast\;\partial_\mu P_1 + i\;e\;\phi^\ast\;(\partial_\mu C)\;\phi + i\;P_2\;\partial_\mu\phi
-e\;A_\mu\;(\phi^\ast\;P_1 +P_2\;\phi) =0,\]
\[i\;\phi\;\partial_\mu P_2 - i\;e\;\phi\;(\partial_\mu C)\;\phi^\ast + i\;P_1\;\partial_\mu\phi^\ast
+e\;A_\mu\;( \phi\; P_2 +P_1\;\phi^\ast) =0.\eqno(A.10)\]
It is straightforward to note that the following choices of the secondary fields $(P_1(x), P_2(x))$
satisfy the above relations (because the first three terms  and the coefficient of $e\;A_\mu$
 should vanish separately and independently), namely;
\[P_1 = -\;e\;C\;\phi,      \qquad P_2  = +\; e\;\phi^\ast C.\eqno(A.11)\]
It is pertinent to point out that the $(\pm)$ signs in $P_2$ and $P_1$ are fixed.
The choices of $(\pm)$ signs in (A.11) are further fixed by the requirement $s_b(\bar D_\mu\phi^\ast\;D^\mu\phi) = 0$ 
which amounts to the following restrictions on the superfields:
\[(\partial_\mu - i\;e\; B_\mu^{(b)}(x,\bar\theta))\;\Phi^\ast(x,\bar\theta)\;(\partial^\mu + i\;e\; B^{\mu(b)}(x,\bar\theta))\;\Phi(x,\bar\theta)\]
\[ = (\partial_\mu - i\;e\; A_\mu(x))\;\phi^\ast(x)\;
(\partial^\mu + i\;e\; A^\mu(x))\;\phi(x).\eqno(A.12)\]
The substitutions of expansions from (A.6) and (20) lead to:
\[i\;e^2 A_\mu\;A^\mu (\phi^\ast\;P_1 + P_2\;\phi) + i\;\partial_\mu\phi^\ast \;[\partial^\mu P_1 + e\; (\partial^\mu C)\; \phi]
+ i\; \partial_\mu\phi\; [ \partial^\mu P_2 - e\; (\partial^\mu C)\;\phi^\ast]\]
\[  + \;e\;A_\mu \Big[ \phi^\ast\;\partial^\mu\;P_1 + P_2\;\partial^\mu \phi-(\partial^\mu\phi^\ast)\;P_1 - (\partial^\mu P_2)\;\phi + e\;\phi^\ast\;(\partial^\mu C)\;\phi + e\;\phi^\ast\;(\partial^\mu C)\;\phi\Big].\eqno(A.13)\]
It is crystal clear that the choices of $(P_1, P_2)$ (that have been pointed out in (A.11)) satisfy  the above relationships, too.
Thus, the expressions in (A.11) are precise and perfect. We would like to mention here (without giving all the algebraic details)
that the last two restrictions of (A.9), with the substitutions from (A.6) and (25), lead to the following expressions for the secondary fields:
\[ \bar P_1 = - \; e\;\bar C\; \phi, \qquad \bar P_2 = + e\; \phi^\ast \bar C.\eqno(A.14)\]
The above choices are correct because these can be {\it further} confirmed by the requirement $s_{ab} (\bar D_\mu\phi^\ast\; D^\mu\phi) = 0$
and corresponding restriction on the superfields (as has been done in Eqs. (A.12) and (A.13)).
Hence, we obtain the explicit expressions of the super expansions of the  matter superfields (i.e. super complex scalar fields), 
after the application of (anti-) BRST invariant restrictions (A.7) and (A.8), as
\[\Phi ^{(b)}(x, \bar\theta) = \phi (x) + \bar\theta\; (-\,i\,e\,C\; \phi (x))\equiv \phi (x) + \bar\theta\;(s_b \;\phi),\]    
\[\Phi^{\ast(b)}(x, \bar\theta) = \phi^\ast (x) + \bar\theta\; (+\,i\,e\, \phi^\ast (x)\; C)\equiv \phi^\ast (x) + \bar\theta\;(s_{b} \;\phi^\ast),\] 
\[\Phi ^{(ab)}(x, \theta) = \phi (x) + \theta\; (-\,i\,e\,\bar C \;\phi (x))\equiv \phi (x) + \theta\;(s_{ab} \;\phi),\]
\[\Phi ^{\ast (ab)}(x, \theta) = \phi^\ast (x) + \theta\; (+\,i\,e\, \phi^\ast(x)\;\bar C )\equiv \phi^\ast (x) + \theta\;(s_{ab} \;\phi^\ast),\eqno(A.15)\]
 where the superscripts $(b)$ and $(ab)$ denote the expansions  for the superfields after application of the BRST and anti-BRST
 invariant restrictions (cf. (A.7) and (A.8)).  It is crystal clear that we have derived {\it all} the proper (anti-)BRST 
transformations for {\it all} the fields of our {\it interacting} Abelian 1-form gauge theory where there is a coupling between 
the $U(1)$ gauge field $(A_\mu)$ and matter fields $(\phi, \phi^\ast)$ which are nothing but the complex scalar fields.

We end this the Appendix with the remarks that the {\it final} expressions for the 
nilpotent and conserved (anti-)BRST charges $(Q_{(a)b})$ for the 
Abelian 1-form {\it interacting} gauge theories with Dirac fields $(\psi,\bar\psi)$ and complex scalar fields $(\phi,\phi^\ast)$
are {\it one} and the {\it same} (cf. Eqs. (6) and (7)). Thus, we note that the proof of their nilpotency and absolute anticommutativity
properties would go along the same lines (as far as the (anti-)chiral superfields approach to BRST formalism is concerned).
In other words, the discussions between the Eqs. (26) and (35) would be {\it same} for the {\it interacting} Abelian 1-form gauge theory with 
complex scalar fields. Furthermore, invariance of the Lagrangian density (A.1) would be {\it same} as discussed between the Eqs. (36)
and (40) within the framework of augmented version of (anti-)chiral superfields approach to BRST formalism.
Thus, we shall {\it not} discuss these aspects, once again, here.\\

\begin{center}
{\bf Appendix B: On Full Super-Expansion and Absolute Anticommutativity}\\
\end{center}
\noindent

In order to corroborate and establish the {\it novelty} of our present investigation with (anti-) chiral superfields and their 
super expansions, we show here that the absolute anticommutativity of the (anti-)BRST symmetries and corresponding conserved 
charges is very {\it natural} and automatic when we consider the {\it full} super expansions of the superfields. 
For instance, let us take a generic superfield $\Sigma  (x, \theta , \bar\theta)$, defined on a general $(D, 2)$-dimensional supermanifold,
with the following super expansion
\[\Sigma (x, \theta , \bar\theta) = \sigma (x) + \theta\;\bar M(x) + \bar\theta\;M(x) + i\;\theta\;\bar\theta\;N(x),\eqno(B.1)\] 
where $\sigma (x)$ is an ordinary D-dimensional field defined on a given D-dimensional ordinary Minkowskian  spacetime manifold  for a given BRST invariant gauge theory.
If $\sigma (x)$ were fermionic, the superfield $\Sigma  (x, \theta , \bar\theta)$ would {\it also} be fermionic 
which implies that the pair $(M(x), \bar M(x))$ would be bosonic and $N(x)$ would be fermionic. On the other hand, if $\sigma (x)$ were bosonic,
 the corresponding superfield $\Sigma  (x, \theta , \bar\theta)$ would be bosonic, too, 
and the pair $(M(x), \bar M(x))$ would be fermionic and $N(x)$ would be bosonic. These conclusions are straightforward  due to 
the fact that the Grassmannian variables $(\theta, \bar\theta)$ are fermionic  (i.e. $\theta^2 =\bar\theta^2 = 0, \theta\bar\theta+\bar\theta\theta = 0$) in nature.
It is straightforward to note that the following are true, namely;
\[\frac {\partial}{\partial\theta}\frac{\partial}{\partial\bar\theta}\Sigma (x,\theta,\bar\theta) = - \;i\; N(x),\qquad\quad \frac {\partial}{\partial\bar\theta}\frac{\partial}{\partial\theta}\Sigma (x,\theta,\bar\theta) = +\; i\; N(x).\eqno(B.2)\]
Taking into account the fact that $\partial_{\theta}\leftrightarrow s_{ab}$ and $\partial_{\bar\theta}\leftrightarrow s_b$,
we observe that the relationship $(\partial_\theta \partial_{\bar\theta}+\partial_{\bar\theta}\partial_\theta)\;\Sigma (x,\theta,\bar\theta) = 0 $, 
in its operator form, implies  the following:
\[(\partial_\theta\; \partial_{\bar\theta}+\partial_{\bar\theta}\;\partial_\theta) = 0\quad \Longleftrightarrow \quad s_{ab}\;s_b + s_b\; s_{ab} = 0.\eqno (B.3)\]
Thus, it is crystal clear that the (anti-)BRST symmetries $s_{(a)b}$ are absolutely anticommuting (i.e. $s_{ab}s_b + s_b s_{ab} = 0$) in nature 
within the framework of superfield approach to BRST formalism if we take into account the {\it full} super expansions of the 
superfields along the $(\theta, \bar\theta)$-directions of (D, 2)-dimensional supermanifold. The relationship (B.3) is {\it not}
guaranteed  when we take into account {\it only} the (anti-)chiral super expansions of the superfields (as is the case in our present
investigation) and in the discussion of $\mathcal {N} = 2$ SUSY quantum mechanical models [16-19] where the SUSY transformations are nilpotent 
(i.e. fermionic) in nature {\it but} not absolutely anticommuting (i.e. $not$ linearly independent of each-other).

We end this Appendix with the remark that the generators of the infinitesimal
 and continuous (anti-)BRST symmetry transformations 
are the conserved (anti-) BRST charges and {\it both} are connected (cf. Sec. 2) by the 
relationship (8) . As a consequence,
we infer that the conserved (anti-)BRST charges $(Q_{(a)b})$ would {\it also} 
obey the absolute anticommutativity property
(i.e. $Q_b Q_{ab}+ Q_{ab} Q_b = 0$) when the {\it full} super expansions 
of the superfields are taken into account along the
$(\theta, \bar\theta)$-directions of (D, 2)-dimensional supermanifold. 
The {\it novel} observation, in our present investigation, is the 
fact that the absolute anticommutativity property of the (anti-)BRST 
charges has been proven  {\it despite} the fact that we have 
considered {\it only} the (anti-)chiral super expansions
 of the (anti-)chiral superfields (defined on  the (D, 1)-dimensional super-submanifolds
of the {\it general} (D, 2)-dimensional supermanifold on which our D-dimensional {\it ordinary} gauge theory is generalized).

\end{document}